# An algorithm for the *k*-error linear complexity of a sequence with period $2p^n$ over GF(q)

Jianqin Zhou
(Dept. of Computer Science, Anhui University of Technology, Ma'anshan 243002, China)
(E-mail: zhou9@yahoo.com)
Xirong Xu
(Dept. of Mathematics, Univ. of Science and Technology of China, Hefei 230026, China)
(E-mail: xirongxu@ustc.edu.cn)

**Abstract--** The union cost is used, so that an efficient algorithm for computing the *k*-error linear complexity of a sequence with period $2p^n$ over GF(q) is presented, where p and q are odd primes, and q is a primitive root of modulo $p^2$.

**Index Terms--** Periodic sequence; linear complexity; *k*-error linear complexity

## I. INTRODUCTION

The linear complexity (LC) of a sequence has been used as a convenient measure of the randomness of a sequence. However, the LC has such an instability as an extreme change. The *k*-error LC (*k*-LC) of a periodic sequence was defined by Stamp and Martin in [4] as the smallest LC that can be obtained when any *k* or fewer of the symbols of the sequence are changed within one period. The *k*-LC is very effective for reducing the instability of the LC caused by symbol substitutions.

Unfortunately an effective algorithm for computing the *k*-LC has been known only for sequences over GF(2) with period $2^n$ (the Stamp-Martin algorithm) in [4]. An alternative derivation of the Stamp-Martin algorithm was given in [3] for computing the *k*-error linear complexity of sequences over $GF(p^m)$ with period $p^n$, p a prime.

This paper gives a complete description of the algorithm for the *k*-LC of sequences over GF(q) with period $2p^n$, where p and q are odd primes, and q is a primitive root of modulo $p^2$. The algorithm is derived from the Wei-Xiao-Chen algorithm in [5] for the linear complexity of sequences over GF(q) with period $2p^n$ and by using the union cost different from that used in the Stamp-Martin algorithm for sequences over GF(2) with period $2^n$. It is shown that both of the logic of our algorithm and its description are rather simple.

## II. THE WEI-XIAO-CHEN ALGORITHM

In this paper we will consider sequences over GF(q) with period $2p^n$, where p and q are odd primes, and q is a primitive root of modulo $p^2$.

Let $s=(a_0, a_1, \cdots)$ be a sequence with period $N=2p^n$ over GF(q), $l>0$, $A_i = (a_{(i-1)l}, a_{(i-1)l+1}, \cdots, a_{il-1})$, i=1,2,$\cdots$,2p. It is easy to prove the following lemmas.

**Lemma 1:** If $\begin{cases} A_{p+1} + A_1 = A_{p+2} + A_2 = \mathbf{L} = A_{2p} + A_p \\ A_{p+1} - A_1 = (-1)^{i+1}(A_{p+i} - A_i), i=1,2,\mathbf{L},p \end{cases}$, then $\begin{cases} A_1 = A_3 = \mathbf{L} = A_p = A_{p+2} = \mathbf{L} = A_{2p-1} \\ A_{p+1} = A_{p+3} = \mathbf{L} = A_{2p} = A_2 = \mathbf{L} = A_{p-1} \end{cases}$,

hence $(A_1, A_2) = (A_1, A_{p+1})$.

**Lemma 2:** If $A_{p+1} + A_1 = A_{p+2} + A_2 = \mathbf{L} = A_{2p} + A_p$, then $A_i - A_{i+1} = -(A_{p+i} - A_{p+i+1})$, $i=2,4,\mathbf{L},p-1$, hence

$(\sum_{i=1}^{p}(-1)^{i+1}A_i, \sum_{i=1}^{p}(-1)^{i+1}A_{i+1}) = (\sum_{i=1}^{p}(-1)^{i+1}A_i, \sum_{i=1}^{p}(-1)^{i+1}A_{p+i})$.

**Lemma 3:** If $A_{p+1} - A_1 = (-1)^{i+1}(A_{p+i} - A_i)$, $i=1,2,\mathbf{L},p$, then $A_i + A_{i+1} = A_{p+i} + A_{p+i+1}$, $i=2,4,\mathbf{L},p-1$, hence

$(\sum_{i=1}^{p} A_i, \sum_{i=1}^{p} A_{i+1}) = (\sum_{i=1}^{p} A_i, \sum_{i=1}^{p} A_{p+i})$.

With the above lemmas, the Wei-Xiao-Chen algorithm in [5] can be changed to the algorithm 1 in Fig.1.

Let $s=(a_0, a_1, \cdots)$ be a sequence with period $N=2p^n$ over GF(q), where p and q are odd primes, and q is a primitive root of modulo $p^2$, and let $s^N=(a_0, a_1, \cdots, a_{N-1})$ be the first period of s.

    a= $s^N$; $l= p^n$; c=0;
    while $l > 1$ do
        $l = l/p$; $A_i = (a_{(i-1)l}, a_{(i-1)l+1}, \cdots, a_{il-1})$, for i=1,2,$\cdots$,2p;
        if $A_{p+1} + A_1 = A_{p+2} + A_2 = \mathbf{L} = A_{2p} + A_p$ then
            if $A_{p+1} - A_1 = (-1)^{i+1}(A_{p+i} - A_i)$ for $i=1,2,\mathbf{L},p$ then
                a=$(A_1, A_{p+1})$;
            else

$$c=c+(p-1)l;\ a=(\sum_{i=1}^{p}(-1)^{i+1}A_i, \sum_{i=1}^{p}(-1)^{i+1}A_{p+i});$$

   end if
  else
   if $A_{p+1}-A_1=(-1)^{i+1}(A_{p+i}-A_i)$ for $i=1,2,\mathbf{L},p$ then

$$c=c+(p-1)\ l;\ a=(\sum_{i=1}^{p}A_i, \sum_{i=1}^{p}A_{p+i});$$

   else

$$c=c+2(p-1)\ l;\ a=(\sum_{i=1}^{p}A_{2i-1}, \sum_{i=1}^{p}A_{2i});$$

   end if
  end if
 end while
 if a $\neq$ (0,0) then
  if $a_0=a_1$ then
   c=c+1
  else
   if $a_0+a_1=0$ then
    c=c+1
   else
    c=c+2
   end if
  end if
 end if

Fig.1. Algorithm 1, computing the linear complexity of a sequence with period $2p^n$ over GF(q)

## III. A $k$-ERROR LINEAR COMPLEXITY ALGORITHM

The $k$-LC of a sequence $s=(a_0, a_1, \cdots)$ over GF(q) with period $N=2p^n$ is defined as

 $k$-LC(s)=min{LC(s+e) |$w_H$(e)$\leq k$}

where $e=(e_0, e_1, \cdots)$ is an error sequence over GF(q) with period $N$ and $w_H(e)$ is the Hamming weight of the first $N$-tuple, ($e_0, e_1, \cdots, e_{N-1}$), of e, i.e., the number of nonzero $e_j$'s. If we have no effective algorithm for computing the $k$-LC, we must repeatedly apply the Wei-Xiao-Chen algorithm at the worst case

$$\sum_{i=0}^{k}(q-1)^i\binom{N}{i} \quad (1)$$

times to the sequences (s+e)'s with all the possible e's having Hamming weight $\leq k$. However, (1) becomes very large even for moderate $N$ and $k$.

In order to compute the $k$-LC of s, we must try to force $A_{p+1}+A_1=A_{p+2}+A_2=\mathbf{L}=A_{2p}+A_p$ and $A_{p+1}-A_1=(-1)^{i+1}(A_{p+i}-A_i)$ for $i=1,2,\mathbf{L},p$, in the Wei-Xiao-Chen algorithm under the condition that the minimum number of changes in the original $s^N$ is less than or equal to $k$. This logic is the same as that used in the Stamp-Martin algorithm in [4].

In [3,4], cost[i] is intended to measure the cost-in terms of the minimum number of changes required in the original sequence s-of changing the current element $a_i$ without disturbing the results of any previous steps. Due to the condition that $A_{p+1}+A_1=A_{p+2}+A_2=\mathbf{L}=A_{2p}+A_p$ and $A_{p+1}-A_1=(-1)^{i+1}(A_{p+i}-A_i)$ for $i=1,2,\mathbf{L},p$, the cost of changing the element $a_i$ and the cost of changing the element $a_{l+i}$ are interrelated. Thus the union cost is used to measure the cost of changing $a_i$ and $a_{l+i}$ at the same time.

In the Stamp-Martin algorithm, only the cost of changing the current element is measured. In fact, the cost of maintaining the current element unchanged is sometimes not zero. In our algorithm, cost[i,i+$l$,$h_0$,$h_1$]$^{(2l)}$ is the minimum number of changes required in the original sequence s to change the current element $a_i$ to $h_0$ and the current element $a_{l+i}$ to $h_1$, where $h_0=0,1,\cdots,q-1$, $h_1=0,1,\cdots,q-1$, and $2l$ is the number of current elements. When $l=p^m$, the initial value of cost[i,i+$l$, $a_i$,$a_{l+i}$]$^{(2l)}$ is 0, the initial value of both cost[i,i+$l$, $a_i$,$a_{l+i}+b$]$^{(2l)}$ and cost[i,i+$l$, $a_i+a$,$a_{l+i}$]$^{(2l)}$ is 1, the initial value of cost[i,i+$l$, $a_i + \boldsymbol{a}, a_{l+i} + \boldsymbol{b}]^{(2l)}$ is 2, where i = 0,1,$\cdots$, $l$-1, and $\boldsymbol{a}$ =1,2, $\cdots$,q$-1$, $\boldsymbol{b}$ =1,2, $\cdots$,q$-1$.

Based on algorithm 1, our algorithm for computing the *k*-error linear complexity is written in Fig.2.

Let s=($a_0$, $a_1$,$\cdots$) be a sequence with period $N=2p^n$ over GF(q), where p and q are odd primes, and q is a primitive root of modulo $p^2$, and let $s^N$=($a_0$, $a_1$,$\cdots$, $a_{N-1}$) be the first period of s.

a= $s^N$; $l$= $p^n$ ; c=0;

cost[i,i+$l$, $a_i$,$a_{l+i}$]$^{(2l)}$=0, cost[i,i+$l$, $a_i$,$a_{l+i}$ + $\boldsymbol{a}$ ]$^{(2l)}$ =1, cost[i,i+$l$, $a_i$ + $\boldsymbol{a}$ ,$a_{l+i}$]$^{(2l)}$=1, cost[i,i+$l$, $a_i$ + $\boldsymbol{a}$ ,$a_{l+i}$ + $\boldsymbol{b}$ ]$^{(2l)}$=2, for i = 0,1,$\cdots$, $l-1$, $\boldsymbol{a}$ =1,2, $\cdots$,q$-1$, $\boldsymbol{b}$ =1,2, $\cdots$,q$-1$;

while $l > 1$ do

$l = l/p$ ; $A_i$ = ($a_{(i-1)l}$, $a_{(i-1)l+1}$,$\cdots$, $a_{il-1}$), for i=1,2,$\cdots$,2p;

( $B_1$, $B_2$,$\cdots$, $B_p$ )=( $A_1+A_{p+1}$, $A_2 +A_{p+2}$,$\cdots$, $A_p +A_{2p}$ ),

bcost[i, h]$^{(pl)}$=min{cost[i, i+ p$l$, $d_1$, $d_2$]$^{(2pl)}$|$d_1+d_2$=h}, for h=0,1,$\cdots$,q$-1$, i=0,1,$\cdots$, p$l$-1;

bcost[i, h]$^{(l)}$=$\sum_{j=0}^{p-1}$ bcost[i+j$l$, h]$^{(pl)}$ , for h=0,1,$\cdots$,q$-1$, i=0,1,$\cdots$, $l$-1;

$T_B$=$\sum_{i=0}^{l-1} \min_{0 \le h < q}$ {bcost[i, h]$^{(l)}$};

if $T_B \le k$ then

cost[i, i+$l$, $h_0$, $h_1$]$^{(2l)}$= $\sum_{j=0}^{(p-1)/2}$ cost[i+2$l$j, i+p$l$ +2$l$j, $h_0$, $h_1$]$^{(2pl)}$

+ $\sum_{j=0}^{(p-1)/2-1}$ cost[i+(2j+1)$l$, i+p$l$ +(2j+1)$l$, $h_1$,$h_0$]$^{(2pl)}$, for $h_0$=0,1,$\cdots$,q$-1$, $h_1$=0,1,$\cdots$,q$-1$, i=0,1,$\cdots$,$l$-1;

$T_C$=$\sum_{i=0}^{l-1} \min_{0 \le h_0 < q, 0 \le h_1 < q}$ {cost[i, i+$l$, $h_0$, $h_1$]$^{(2l)}$};

If $T_C \le k$ then

a= ($A_1$,$A_{p+1}$);

else

c=c+(p-1)$l$; a=($\sum_{i=1}^{p} (-1)^{i+1} A_i$, $\sum_{i=1}^{p} (-1)^{i+1} A_{p+i}$) ;

cost[i, i+$l$, $h_0$, $h_1$]$^{(2l)}$=min{ $\sum_{j=0}^{p-1}$ cost[i+$l$j, i+p$l$ +$l$j, $h^0_{i+lj}$, $h^1_{i+lj}$ ]$^{(2pl)}$

| $h^0_{i+lj}$ + $h^1_{i+lj}$ = $h^0_i$ + $h^1_i$, for j=1,2,$\cdots$,p-1, $\sum_{j=0}^{p-1} (-1)^j h^0_{i+lj}$ =$h_0$, $\sum_{j=0}^{p-1} (-1)^j h^1_{i+lj}$ =$h_1$ },

for $h_0$=0,1,$\cdots$,q$-1$, $h_1$=0,1,$\cdots$,q$-1$, i=0,1,$\cdots$, $l$-1;

end if

else

($D_1$, $D_2$,$\cdots$, $D_p$ )=( $A_{p+1}- A_1$, $-(A_{p+2} - A_2$ ),$\cdots$, $A_{2p} -A_p$ );

dcost[i+j$l$, h]$^{(pl)}$=min{cost[i+j$l$, i+j$l$+p$l$, $h^0_{i+lj}$, $h^1_{i+lj}$ ]$^{(2pl)}$| $(-1)^j$ ($h^1_{i+lj} - h^0_{i+lj}$ )=h},

for h=0,1,$\cdots$,q$-1$, i=0,1,$\cdots$, $l$-1, j=0,1,$\cdots$, p-1;

dcost[i, h]$^{(l)}$=$\sum_{j=0}^{p-1}$ dcost[i+j$l$, h]$^{(pl)}$ , for h=0,1,$\cdots$,q$-1$, i=0,1,$\cdots$, $l$-1;

$T_D$=$\sum_{i=0}^{l-1} \min_{0 \le h < q}$ {dcost[i, h]$^{(l)}$};

if $T_D \le k$ then

c=c+(p-1)$l$; a=($\sum_{i=1}^{p} A_i$, $\sum_{i=1}^{p} A_{p+i}$) ;

$$\text{cost}[i, i+l, h_0, h_1]^{(2l)} = \min \{ \sum_{j=0}^{p-1} \text{cost}[i+lj, i+pl+lj, h^0_{i+lj}, h^1_{i+lj}]^{(2pl)}$$

$$| \ (-1)^j (h^1_{i+lj} - h^0_{i+lj}) = h^1_i - h^0_i, \text{ for } j=1,2,\cdots,p-1, \ \sum_{j=0}^{p-1} h^0_{i+lj} = h_0, \ \sum_{j=0}^{p-1} h^1_{i+lj} = h_1 \},$$

for $h_0 = 0, 1, \cdots, q-1$, $h_1 = 0, 1, \cdots, q-1$, $i = 0, 1, \cdots, l-1$;

　　else

$$c = c + 2(p-1)\,l; \ a = (\sum_{i=1}^{p} A_{2i-1}, \sum_{i=1}^{p} A_{2i});$$

$$\text{cost}[i, i+l, h_0, h_1]^{(2l)} = \min \{ \sum_{j=0}^{(p-1)/2} \text{cost}[i+2lj, i+pl+2lj, h^0_{i+2lj}, h^1_{i+2lj}]^{(2pl)}$$

$$+ \sum_{j=0}^{(p-1)/2-1} \text{cost}[i+(2j+1)l, i+pl+(2j+1)l, h^1_{i+(2j+1)l}, h^0_{i+(2j+1)l}]^{(2pl)} | \ \sum_{j=0}^{p-1} h^0_{i+lj} = h_0, \ \sum_{j=0}^{p-1} h^1_{i+lj} = h_1 \},$$

for $h_0 = 0, 1, \cdots, q-1$, $h_1 = 0, 1, \cdots, q-1$, $i = 0, 1, \cdots, l-1$;

　　　　end if

　　end if

end while

if $\text{cost}[0, 1, 0, 0]^{(2)} > k$ then

　if $\min_{0 \le h < q} \{ \text{cost}[0, 1, h, h]^{(2)} \} \le k$　then

　　$c = c+1$

　else

　　if $\min_{h_0 + h_1 = 0} \{ \text{cost}[0, 1, h_0, h_1]^{(2)} \} \le k$　then

　　　$c = c+1$

　　else

　　　$c = c+2$

　　end if

　end if

end if

Fig.2. Algorithm 2, computing the *k*-error linear complexity of a sequence with period $2p^n$ over GF(q)

This new algorithm reduces to the Wei-Xiao-Chen algorithm [5] in the case $k = 0$. The validity of our algorithm can be shown by using the following propositions.

**Proposition 1.** At the jth($j \le n$) step, we may prevent $(p-1)p^{n-j}$ or $2(p-1)p^{n-j}$ from being added to c, and the total of all remaining possible additions is only $2p^{n-j}$.

**Proof:**
$$\begin{bmatrix} (p-1)p^{n-1} & (p-1)p^{n-2} & \mathbf{L} & (p-1)p & p-1 & 1 \\ (p-1)p^{n-1} & (p-1)p^{n-2} & \mathbf{L} & (p-1)p & p-1 & 1 \end{bmatrix}$$

Here the jth column of the matrix represents the possible additions to c at the jth step.

The total of all remaining possible additions is $2(p-1)p^{m-j-1} + 2(p-1)p^{m-j-2} + \cdots + 2(p-1) + 2 = 2p^{m-j}$.

**Proposition 2.** $\sum_{i=0}^{l-1} \min_{0 \le h_0 < q, 0 \le h_1 < q} \{ \text{cost}[i, i+l, h_0, h_1]^{(2l)} \} \le k$, $l = p^n, \cdots, p, 1$

**Proof:** When $l = p^n$, it is easy to show that $\sum_{i=0}^{l-1} \min_{0 \le h_0 < q, 0 \le h_1 < q} \{ \text{cost}[i, i+l, h_0, h_1]^{(2l)} \} = 0$.

If $T_C \le k$ at the jth step, we have $\sum_{i=0}^{l-1} \min_{0 \le h_0 < q, 0 \le h_1 < q} \{ \text{cost}[i, i+l, h_0, h_1]^{(2l)} \} = T_C \le k$.

If $T_B \le k$ and $T_C > k$ at the jth step, we have $\sum_{i=0}^{l-1} \min_{0 \le h_0 < q, 0 \le h_1 < q} \{ \text{cost}[i, i+l, h_0, h_1]^{(2l)} \} = T_B \le k$.

If $T_D \le k$ and $T_B > k$ at the jth step, we have $\min_{0 \le h_0 < q, 0 \le h_1 < q} \{ \text{cost}[i, i+l, h_0, h_1]^{(2l)} \} = T_D \le k$.

If $T_D > k$ and $T_B > k$ at the jth step, we have

$$\sum_{i=0}^{l-1} \min_{0\leq h_0<q, 0\leq h_1<q} \{ \text{cost}[i, i+l, h_0, h_1]^{(2l)}\} = \sum_{i=0}^{pl-1} \min_{0\leq h_0<q, 0\leq h_1<q} \{ \text{cost}[i, i+pl, h_0, h_1]^{(2pl)}\} \leq k.$$ The proof is completed.

## IV. NUMERICAL EXAMPLE

**Example 1.** Let $s$ be a sequence over GF(3) with period $N=2 \cdot 5^2$ whose one period is
$s^N$= 1210100012 1202102202 1122112121 2121012101 2102110100. We will compute the 5-error linear complexity of $s$.
**Initial values:** a= $s^N$; $l= 5^2$; c=0;

$$(\text{cost}[i,i+l,h_0,h_1]^{(2l)})=\begin{pmatrix} 2 & 2 & 2 & 1 & 2, & 1 & 1 & 1 & 2 & 1, & 2 & 2 & 1 & 1 & 2, & 1 & 2 & 1 & 1 & 2, & 2 & 1 & 2 & 1 & 1 \\ 1 & 2 & 1 & 1 & 1 & 1 & 0 & 1 & 1 & 2 & 1 & 2 & 0 & 2 & 1 & 1 & 1 & 2 & 1 & 1 & 1 & 2 & 1 & 2 & 2 \\ 2 & 1 & 2 & 0 & 2 & 0 & 1 & 0 & 2 & 2 & 2 & 1 & 1 & 2 & 2 & 0 & 2 & 2 & 0 & 2 & 2 & 2 & 2 & 2 & 2 \\ 1 & 2 & 1 & 2 & 1 & 2 & 2 & 2 & 1 & 1 & 1 & 2 & 2 & 1 & 1 & 2 & 2 & 1 & 2 & 2 & 1 & 0 & 2 & 1 & 0 \\ 0 & 2 & 0 & 2 & 0 & 2 & 1 & 2 & 0 & 2 & 0 & 2 & 1 & 2 & 0 & 2 & 1 & 2 & 2 & 1 & 0 & 1 & 1 & 2 & 1 \\ 1 & 1 & 1 & 1 & 1 & 1 & 2 & 1 & 1 & 2 & 1 & 1 & 2 & 2 & 1 & 1 & 2 & 2 & 1 & 2 & 1 & 1 & 2 & 2 & 1 \\ 2 & 1 & 2 & 2 & 2 & 2 & 2 & 2 & 0 & 2 & 1 & 2 & 0 & 2 & 2 & 1 & 0 & 2 & 1 & 2 & 1 & 1 & 0 & 1 \\ 1 & 1 & 1 & 2 & 1 & 2 & 1 & 2 & 1 & 1 & 1 & 1 & 1 & 1 & 2 & 0 & 1 & 2 & 0 & 1 & 2 & 0 & 1 & 2 \\ 2 & 0 & 2 & 1 & 2 & 1 & 2 & 1 & 2 & 1 & 2 & 0 & 2 & 1 & 2 & 1 & 1 & 1 & 1 & 2 & 2 & 1 & 1 & 2 \end{pmatrix},$$

where the ith($0\leq i<l$) column of the matrix represents $\text{cost}[i,i+l,h_0,h_1]^{(2l)}$, $h_0=0,1,\cdots,q-1$, $h_1=0,1,\cdots,q-1$.

**Step 1.** $l=5$,    $A_1$=12101, $A_2$=00012, $A_3$=12021, $A_4$=02202, $A_5$=11221, $A_6$=12121, $A_7$=21210, $A_8$=12101, $A_9$=21021, $A_{10}$=10100

$$(\text{bcost}[i,h]^{(pl)})=\begin{pmatrix} 1 & 1 & 1 & 1 & 1, & 1 & 1 & 1 & 1 & 1, & 1 & 1 & 1 & 1 & 1, & 1 & 0 & 1 & 1 & 0, & 1 & 1 & 0 & 1 & 1 \\ 1 & 0 & 1 & 1 & 1 & 1 & 0 & 1 & 1 & 1 & 1 & 0 & 0 & 1 & 1 & 1 & 1 & 1 & 1 & 1 & 1 & 0 & 1 & 1 & 0 \\ 0 & 1 & 0 & 0 & 0 & 0 & 1 & 0 & 0 & 0 & 0 & 1 & 1 & 0 & 0 & 0 & 1 & 0 & 0 & 1 & 0 & 1 & 1 & 0 & 1 \end{pmatrix}$$

$$(\text{bcost}[i,h]^{(l)})=\begin{pmatrix} 5 & 4 & 4 & 5 & 4 \\ 5 & 1 & 4 & 5 & 4 \\ 0 & 5 & 2 & 0 & 2 \end{pmatrix}, \quad T_B=5=k. \quad (\text{cost}[i, i+l, h_0, h_1]^{(2l)})=\begin{pmatrix} 8 & 8 & 7 & 6 & 8 \\ 7 & 10 & 5 & 8 & 7 \\ 10 & 7 & 7 & 8 & 7 \\ 5 & 5 & 8 & 6 & 5 \\ 4 & 7 & 6 & 8 & 4 \\ 7 & 4 & 8 & 8 & 4 \\ 6 & 6 & 7 & 4 & 9 \\ 5 & 8 & 5 & 6 & 8 \\ 8 & 5 & 7 & 6 & 8 \end{pmatrix}$$

$T_C$=21>$k$, so c=20.   $(\text{cost}[i, i+l, h_0, h_1]^{(2l)})=\begin{pmatrix} 5 & 4 & 4 & 5 & 4 \\ 5 & 1 & 4 & 5 & 4 \\ 0 & 5 & 2 & 0 & 2 \\ 5 & 2 & 4 & 5 & 4 \\ 2 & 5 & 2 & 2 & 2 \\ 5 & 4 & 4 & 5 & 4 \\ 2 & 5 & 2 & 2 & 2 \\ 5 & 4 & 4 & 5 & 4 \\ 5 & 1 & 4 & 5 & 4 \end{pmatrix}$

**Step 2.** $l=1$,    $A_1$=0, $A_2$=0, $A_3$=1, $A_4$=0, $A_5$=2, $A_6$=2, $A_7$=2, $A_8$=1, $A_9$=2, $A_{10}$=1.

$(\text{bcost}[i,h]^{(pl)})=\begin{pmatrix} 5 & 4 & 4 & 5 & 4 \\ 5 & 1 & 4 & 5 & 4 \\ 0 & 5 & 2 & 0 & 2 \end{pmatrix}$, $(\text{bcost}[i,h]^{(l)})=\begin{pmatrix} 22 \\ 19 \\ 9 \end{pmatrix}$, $T_B$=9>$k$.

$$(\text{dcost}[i,h])^{(pl)} = \begin{pmatrix} 2 & 1 & 2 & 2 & 2 \\ 2 & 2 & 2 & 0 & 2 \\ 0 & 1 & 2 & 2 & 2 \end{pmatrix}, (\text{dcost}[i,h])^{(l)} = \begin{pmatrix} 9 \\ 8 \\ 7 \end{pmatrix}. T_D = 7 > k, \text{ so } c = 20 + 8 = 28. (\text{cost}[i, i+l, h_0, h_1])^{(2l)} = \begin{pmatrix} 5 \\ 7 \\ 7 \\ 7 \\ 7 \\ 7 \\ 5 \\ 7 \\ 5 \\ 7 \end{pmatrix}.$$

**Step 3.** $a=(1,1)$. Since $\text{cost}[0, 1, 0, 0]^{(2)} = 5 = k$, therefore $c=28$.
Finally the 5-error linear complexity is 28.

## V. CONCLUSION

First, we optimize the structure of the Wei-Xiao-Chen algorithm in [5] for the linear complexity of sequences over GF(q) with period $N = 2p^n$, where p and q are odd primes, and q is a primitive root (mod $p^2$).

Second, we presented an algorithm for determining the $k$-error linear complexity of a sequence with period $N = 2p^n$ over GF(q), where p and q are odd primes, and q is a primitive root (mod $p^2$). The algorithm is derived from the Wei-Xiao-Chen algorithm for the linear complexity of sequences over GF(q) with period $2p^n$ and by using the union cost different from that used in the Stamp-Martin algorithm for sequences over GF(2) with period $2^n$. The algorithm reduces to the Wei-Xiao-Chen algorithm in the case $k = 0$.

## REFERENCES

[1] C. Ding, G. Xiao, W. Shan, The Stability Theory of Stream Ciphers. Lecture Notes in Computer Science Vol. 561. Berlin/ Heidelberg, Germany: Springer-Verlag, 1991.
[2] R. A. Games, A. H. Chan, "A fast algorithm for determining the complexity of a pseudo-random sequence with period $2^n$". IEEE Trans. Inform. Theory, vol. IT-29, pp. 144-146, Jan.1983.
[3] T. Kaida, S. Uehara, K. Imamura, "An algorithm for the $k$-error linear complexity of sequences over GF($p^m$) with period $p^n$, p a prime". Information and Computation,1999,151(1):134 -147.
[4] M. Stamp, C. F. Martin, "An algorithm for the $k$-error linear complexity of binary sequences with period $2^n$". IEEE Trans. Inform. Theory, vol.39, pp. 1389-1401, July 1993.
[5] S. Wei, G. Xiao, Z.Chen, "A fast algorithm for determining the minimal polynomial of a sequence with period $2p^n$ over GF(q)", IEEE Trans. Inform. Theory, vol.48, pp. 2754-2758, Oct.2002.
[6] G. Xiao, S. Wei, K. Y. Lam, and K. Imamura, "A fast algorithm for determining the linear complexity of a sequence with period $p^n$ over GF(q)", IEEE Trans. Inform. Theory, vol.46, pp. 2203-2206, Sept.2000.

RESUME OF JIANQIN ZHOU

-------------------------------------------------------------------------------------------------
Education:
1986-1989        Statistics Department, Fudan University,
                 Shanghai, P.R.China, Master of Science in Statistics.
1979-1983        Mathematics Department, East China Normal University,
                 Shanghai, P.R.China, Bachelor of Science in Mathematics
-------------------------------------------------------------------------------------------------
Career Summary:
        Professor, Department of Computer Science,
        Anhui University of Technology.
        Published more than thirty five papers, one paper
        proved a conjecture posed by famous mathematician
        Paul Erdos et al. Research interests include theoretical
        computer science, combinatorics and algorithm.
-------------------------------------------------------------------------------------------------